\documentclass[aps,10pt,prl,reprint,superscriptaddress,showpacs,floatfix,longbibliography]{revtex4-1}
\usepackage{mathrsfs,braket}
\usepackage{amssymb, amsbsy, amsmath, latexsym, dsfont, array, layout, graphicx,mathrsfs,color,bm}
\usepackage{xcolor}
\usepackage{physics}
\usepackage[colorlinks=true,citecolor=blue,urlcolor=blue]{hyperref}

\usepackage{hyphenat}
\usepackage{lipsum}

\newcommand{\one}{\mathds{1}}
\newcommand{\ketbrad}[1]{\left|{#1}\rangle\!\langle{#1}\right|}

\begin{document}

\title{Observation of Braid-Protected Unpaired Exceptional Points}


\author{Kunkun Wang}\thanks{These authors contributed equally to this work}
\affiliation{School of Physics and Optoelectronic Engineering, Anhui University, Hefei 230601, China}
\author{J. Lukas K. König}\thanks{These authors contributed equally to this work}
\affiliation{Department of Physics, Stockholm University, AlbaNova University Center, 106 91 Stockholm, Sweden}
\author{Kang Yang}
\affiliation{Dahlem Center for Complex Quantum Systems and Fachbereich Physik, Freie Universit\"at Berlin, 14195 Berlin, Germany}
\author{Lei Xiao}
\affiliation{School of Physics, Southeast University, Nanjing 211189, China}
\author{Wei Yi}
\affiliation{Laboratory of Quantum Information, University of Science and Technology of China, Hefei 230026, China}
\affiliation{Anhui Province Key Laboratory of Quantum Network, University of Science and Technology of China, Hefei, 230026, China}
\affiliation{CAS Center For Excellence in Quantum Information and Quantum Physics, Hefei 230026, China}
\author{Emil J. Bergholtz}\email{emil.bergholtz@fysik.su.se}
\affiliation{Department of Physics, Stockholm University, AlbaNova University Center, 106 91 Stockholm, Sweden}
\author{Peng Xue}\email{gnep.eux@gmail.com}
\thanks{Present address: Beijing Computational Science Research Center, Beijing 100193, China}
\affiliation{School of Physics, Southeast University, Nanjing 211189, China}

\begin{abstract}
{Spectral degeneracies (dubbed nodal points in momentum space) play fundamental roles in understanding exotic properties of light and matter. In lattice systems, unpaired band-structure degeneracies are subject to well-established no-go (doubling) theorems that universally apply to both closed Hermitian systems and open non-Hermitian systems. However, the non-Abelian braid topology of non-Hermitian multi-band systems provides a loophole to these constraints. Here we successfully leverage this loophole in a non-Hermitian three-band system, implementing an unpaired third-order exceptional point (EP3), which manifests as a non-Abelian monopole. We explicitly demonstrate the intricate braiding topology and non-Abelian, path-dependent, fusion rules underlying the unpaired EP3. The experiment uses a new design of single-photon interferometry, enabling eigenstate and spectral resolutions for multi-band systems with widely tunable parameters. Thus, the union of state-of-the-art experiments, fundamental theory, and everyday concepts such as braids pave the way toward the highly exotic non-Abelian topology unique to non-Hermitian settings.}
\end{abstract}
\maketitle

{\it Introduction.---}
Nodal points, the spectral degeneracies of band structures, often exhibit remarkable and unique properties. In particular, the discovery of Dirac and Weyl semimetals has spurred significant research interest in topologically stable nodal points in Bloch bands~\cite{RevModPhys.90.015001,TZL19,TZZ21}. Their existence is constrained by the famous Nielsen-Ninomiya fermion-doubling theorem~\cite{NIELSEN198120,NIELSEN1981173}, which asserts that, in crystalline systems, nodal points that carry topological charges must appear in pairs. Only very recently have theoretical advancements revealed strategies to circumvent this fundamental constraint.

One caveat of the fermion-doubling theorem is that it requires simply additive charges, such that its traditional form does not apply to non-Abelian charges.
Such non-Abelian charges obey non-commuting statistics. They also follow path-dependent fusion rules, where the result of combining two topological charges depends on the paths that bring them together.
Since non-Abelian fusion outcomes can only be modified by topologically robust braiding operations, non-Abelian charges are considered useful for applications in error-proof information storage and manipulation~\cite{RevModPhys.80.1083}.

Despite their theoretical significance and potential wide applications, implementing non-Abelian nodal points and fusion processes remains challenging.
Dissipative systems promise to close this experimental gap.
Spectral degeneracies in such systems are defective and are commonly referred to as exceptional points (EPs)~\cite{berryPhysicsNonhermitianDegeneracies2004,heissPhysicsExceptionalPoints2012,WLG19}.
These EPs carry such non-Abelian topological charge, and have various practical applications ranging from topological energy transfer in optomechanical systems to the enhancement of sensitivity~\cite{xu2016topological,doppler2016dynamically,hodaei2017enhanced,yanglan2017enhanced,zhong2018winding,naghiloo2019quantum,ashida2020non,bergholtzExceptionalTopologyNonHermitian2021,DFM22,LCA23,TCT24}.

The non-Abelian charge of an EP arises from complex-valued ``energies", which reflect the dissipation of the underlying systems.
These energies braid around each other near the EPs, endowing each of them with a braid topological invariant, which, intriguingly, is non-Abelian for systems of more than two bands~\cite{KT08,wojcikHomotopyCharacterizationNonHermitian2020,liHomotopicalCharacterizationNonHermitian2021,huKnotsNonHermitianBloch2021,wangTopologicalComplexenergyBraiding2021,patilMeasuringKnotNonHermitian2022,yangHomotopySymmetryNonHermitian2023,PhysRevE.87.050101,papNonabelianNatureSystems2018,yang2024non,LLM20,GCD23,LWZ24,WangZ25}.
This opens up the possibility of circumventing the fermion-doubling theorem and constructing braid-protected unpaired EPs in multi-band dissipative systems~\cite{konigBraidprotectedTopologicalBand2023}.
While dissipation is ubiquitous in photonic, mechanical, and electric-circuit systems~\cite{miriExceptionalPointsOptics2019,ozdemir2019parity,PhysRevLett.86.787,dingEmergenceCoalescenceTopological2016,Zhou2018,Maguancong2022,YZL23}, the observation and manipulation of unpaired EPs and non-Abelian phenomena remain elusive, for want of a suitably tunable dissipative system.

In this work, we report the experimental creation and manipulation of a non-Abelian unpaired third-order EP (EP3) and second-order EPs (EP2s) in a single-photon interferometric network (Fig.~\ref{fig:1-setup}).
Distinct from previous studies~\cite{PhysRevLett.130.017201,PhysRevResearch.5.L022050,PhysRevB.109.L041102,guria2024resolving},
for the first time, we control and observe the topological, path-dependent fusions of EP2s, which either produce a non-Abelian unpaired EP3 or open a gap, depending on the topology of their pre-fusion paths while independent of any other EPs.
We further demonstrate that non-Hermitian phase transitions are triggered by moving EP2s along such non-trivial paths.
While the well-established winding numbers fail to detect this transition, we successfully reveal the transition by measuring the non-Abelian braids instead, which clearly highlight the non-Abelian nature of our setup.
In contrast to many-body systems~\cite{sternNonAbelianStatesMatter2010} and Hermitian non-interacting systems~\cite{RNC14,wu2019non,bouhon2020non,PhysRevX.13.021024}, our setup observes non-Abelian braids and fusion without the requirement for fine-tuned interactions or additional symmetries. Instead, the observation of non-Abelian phenomena in our setup only relies on the inherent topological robustness of braids.



\begin{figure}
\centering
\includegraphics[width=\linewidth]{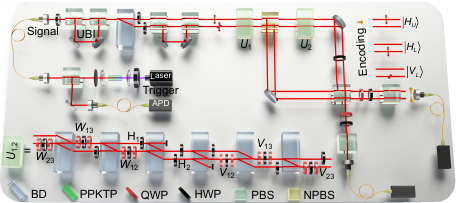}
\caption{
Experimental setup.
Photon pairs (trigger and signal) are generated via type-II spontaneous parametric down-conversion through a periodically poled potassium titanyl phosphate (PPKTP) crystal.
In the state space spanned by horizontally polarised light in the upper spatial mode \(\ket{H_U}\), the lower spatial mode \(\ket{H_L}\), and the vertically polarised lower spatial mode \(\ket{V_L}\),
we prepare the signal photon in a completely mixed qutrit state through two unbalanced interferometer (UBI) setups, consisting of four polarizing beam splitters (PBSs), three half-wave plates (HWPs), and a beam displacer (BD).
We purify this mixed state into one of the eigenstates of the non-Hermitian Hamiltonian $H$ by way of a non-unitary operation $U_1$, realized by six BDs, HWPs and quarter-wave plates (QWPs).
To measure the associated eigenenergy, the photon beams is split by a non-polarizing beam splitter (NPBS): transmitted photons evolve under $U_2$, acquiring a complex phase relative to the reflected path, corresponding to the eigenenergies of $H$.
We measure these eigenenergies interferometrically using avalanche photodiodes (APDs) by detecting coincidences between the signal and trigger photons.
}
\label{fig:1-setup}
\end{figure}

\begin{figure*}
\centering
\includegraphics[width=0.95\textwidth]{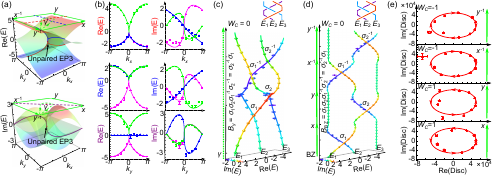}
\caption{
Observation of the Non-Abelian unpaired EP3.
(a) Theoretical results for the real and imaginary parts of the spectrum of $H_M^{\delta=1+2\sqrt{2}}$.
The red dot marks the non-Abelian unpaired EP3.
The solid lines depict the Fermi (i-Fermi) arcs of the real (imaginary) spectral components.
The dashed lines are their projections onto the $k_x$--$k_y$ plane. The path \(\gamma\), with base point at \(k_x=k_y=-\pi\), encircles the unpaired EP3.
(b) The measured eigenenergies along the (i-)Fermi arcs are shown in (a), illustrating the existence of the unpaired EP3 and its dispersion.
(c) The measured braid along \(\gamma\) corresponds to the braid word \(B_\gamma=\sigma_1\sigma_2\sigma_1^{-1}\sigma_2^{-1}\).
(d) The loop following the Brillouin zone (BZ) boundary is homotopy equivalent to \(\gamma\) and thus carries the same braid.
(e) Measured discriminant [see Eq.~\eqref{eq:discriminant}] for the four segments of the loop shown in (c).
Experimental data are represented by dots with error bars indicating statistical uncertainties from Poissonian photon-number fluctuations; theoretical results are denoted by continuous lines.
}
\label{fig:2-monopole}
\end{figure*}

{\it Setup and Protocol.---}
We consider a three-band system on a two-dimensional toric parameter space featuring a single unpaired EP3.
The Hamiltonian
\begin{equation}
\label{eq:monopole-system}
    H_M^\delta =  \mqty(
    \delta-e^{ik_x} & -i(1+e^{ik_x})    & 0 \\
    -i(1+e^{ik_x})  & e^{ik_x}-e^{ik_y} & -i(1+e^{ik_y}) \\
    0               & -i(1+e^{ik_y})    & -\delta +e^{ik_y}
    )
\end{equation}
arises naturally as, for instance, the Bloch Hamiltonian of a lattice-periodic system in two dimensions with three orbitals per unit cell~\cite{konigBraidprotectedTopologicalBand2023}.
The \(2\pi\)-periodic quasimomenta \(k_x\) and \(k_y\) then parameterise the vector \(\mathbf{k}=(k_x,k_y)\) on a Brillouin zone (BZ) torus.
For \(\delta=1+2\sqrt{2}\), the system contains the braid-protected unpaired EP3, a single robust gap-closing, at \(k_x=k_y=0\) [see Fig.~\ref{fig:2-monopole}(a)]. This is forbidden by the fermion-doubling theorem in Hermitian systems, but allowed here by the non-Abelian nature of EPs, which we discuss in the next section.

Experimentally, we encode the three basis states in a subspace of four spatio-polarizational photonic modes and resolve the system spectrum through a three-step protocol (see Fig.~\ref{fig:1-setup}).
First, we prepare the completely mixed qutrit state by subjecting the single photon successively to a pair of unbalanced interferometers (see End Matter).
Second, we implement imaginary-time evolution operator $U_1=e^{f_j(H)\tau_1}$ with the evolution time $\tau_1$. Through appropriate choices of \(f_j(H)\), all eigenstates of the Hamiltonian $H$ damp except for the \(j\)-th one~\cite{MST20,MJE19}.
Third and finally, we obtain the corresponding \(j\)-th complex energy by evolving this state under a real-time evolution $U_2=e^{-i(H-i\Lambda\one)\tau_2}$. The energy can be reconstructed via interferometric measurement between the initial and the evolved states (see End Matter).
Here we introduce the parameter \(\Lambda\), mapping the evolution to a passive operation, thereby bypassing the obstacle of achieving coherent gains in quantum systems.
We then apply this three-step protocol to the Hamiltonian \(H_M^\delta(\mathbf{k})\) in Eq.~\eqref{eq:monopole-system} at representative quasimomenta \(\mathbf{k}=(k_x,k_y)\) throughout the BZ and resolve its complex spectrum.
Experimentally, by adjusting the setting angles of the wave plates $H_1$, $H_2$, $W_{ij}$ and $V_{ij}$, we vary both the quasimomenta \(\mathbf{k}\) and $\delta$ in the Hamiltonian.

Compared to classical systems, such as coupled waveguides~\cite{SST24}, on which non-Hermitian band theory and non-Abelian braids have also been simulated, our method offers several crucial advantages. Specifically, our setup enables direct measurement of complex eigenenergies and full characterization of the instantaneous eigenstates, including their phase information, thereby providing complete access to the braiding dynamics. Our setup also features fully tunable parameters, thus offering more flexibility. Furthermore, by manipulating quantum states, our platform provides a route to the exploration of non-Abelian phenomena in the quantum regime.

{\it Non-Abelian Unpaired EP3.---}
We confirm the presence of the unpaired EP3 at \(k_x=k_y=0\) via spectral measurement, as illustrated in Fig.~\ref{fig:2-monopole}(b).
Only at the unpaired EP3 do all three eigenvalues merge to \(E_{1,2,3}=0\). Near the unpaired EP3, the absolute values of the spectrum generally disperse as $|\delta E|\sim |\delta k|^{1/3}$. Their complex phases strongly depend on the approach direction toward the unpaired EP3.

While the existence of the unpaired EP3 contrasts with conventional fermion-doubling theorems~\cite{NIELSEN198120,NIELSEN1981173,yangFermionDoublingTheorems2021}, we can understand its occurrence by measuring its dispersion, which gives the braid topological invariant protecting it.
We measure the eigenvalues for quasimomenta \( \mathbf{k} \) along the loop $\gamma$, which encloses the unpaired EP3, giving rise to the EP's topological braid charge.
As shown in Fig.~\ref{fig:2-monopole}(c), the complex eigenenergies wind around each other along this loop, forming a non-trivial braid \(B_{\gamma}\). Expressed in terms of simple counterclockwise (clockwise) two-level braiding \(\sigma_i\) (\(\sigma_i^{-1}\)) of energies $E_i$ and $E_{i+1}$, the braid is \(B_{\gamma} =\sigma_1\sigma_2\sigma_1^{-1}\sigma_2^{-1}\). As the braid group is non-Abelian, the EP3 can be regarded as a non-Abelian unpaired EP3.

This braid is a topological invariant of the EP3; for arbitrary continuous perturbations of the loop, it can only change when an EP is crossed.
We verify the unpaired nature of the EP3 by deforming our measurement loop to the BZ boundary~\cite{PhysRevB.106.L161401,PhysRevResearch.4.L022064},
which, for Bloch Hamiltonians, consists of loops along the meridian ($x$) and longitude ($y$) around the BZ torus, following the boundary path \(x y x^{-1} y^{-1}\).
Along this loop, we observe the same braid pattern, \(B_{\text{BZ}} = B_{\gamma}\)~\cite{suppl} (see Fig.~\ref{fig:2-monopole}(d) ).

The boundary loop carries braids \(B_{x,y}\) individually, so the braid \(B_{\text{BZ}}\) splits into four parts,
\begin{equation}\label{eq:non-abelian-invariant}
    B_{\text{BZ}} = B_x B_y B_x^{-1} B_y^{-1}.
\end{equation}
This is at the heart of Fermion doubling; for hypothetical Abelian invariants \(A_{x,y}\), these four parts would commute, resulting in \(A_x A_x^{-1} A_y A_y^{-1}=1\), so topological Abelian unpaired EPs cannot exist.
In comparison, Eq.~\eqref{eq:non-abelian-invariant} illustrates the loophole we utilise here: for non-Abelian braid invariants, this argument fails.
Specifically, for the non-Hermitian three-band model~\eqref{eq:monopole-system}, we identify the boundary invariants \(B_x = \sigma_1, B_y = \sigma_2\) [see Fig.~\ref{fig:2-monopole}(d)].
These do not commute, thus enabling a topologically non-trivial unpaired EP3 protected by the braid invariant \(\sigma_1\sigma_2\sigma_1^{-1}\sigma_2^{-1}\).

In Fig.~\ref{fig:2-monopole}~(e), we illustrate the difference from Abelian two-level systems~\cite{PhysRevE.87.050101,PhysRevLett.118.040401,wangTopologicalComplexenergyBraiding2021,yangFermionDoublingTheorems2021} by showcasing the degree of the unpaired-EP3 braid in the form of the winding number \(W_\mathcal{C}\)~\cite{yangFermionDoublingTheorems2021}, an Abelian invariant that can be derived from a given braid (see End Matter).
As the Abelian part of the eigenenergy braid, \(W_\mathcal{C}\) must vanish when following the boundaries of the BZ~\cite{yangFermionDoublingTheorems2021}. Indeed, while each overcrossing (undercrossing) of the braid contributes a nontrivial winding of the discriminant around the origin, the total winding remains zero, in accordance with the Abelianized sum rule. However, the full non-Abelian braid is not trivial such that the phase is protected and remains nodal under perturbations.

\begin{figure*}
\centering
\includegraphics[width=0.95\textwidth]{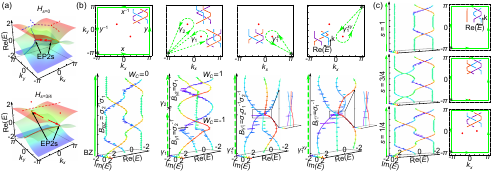}
\caption{
Path-dependent fusion.
(a) Theoretical spectra (real parts) over the BZ, with EP2s marked as red dots.
Top: \(H_M^{\delta=1}=H_{s=0}\).
Bottom: \(H_{s=3/4}\).
(b) Panel 1, Loop along the BZ boundary based at \((k_x,k_y) = (-\pi,-\pi)\) (top) and eigenenergy braid of \(H_M^{\delta=1}\) along this loop (bottom).
Experimental data are presented as points with error bars indicating statistical uncertainties from Poissonian photon-number fluctuations; theoretical predictions are shown as solid lines.
Panel 2: as panel 1.
The two EP2s carry braid invariants \(B_{\gamma_1}=\sigma_2^{-1}\) and \(B_{\gamma_2} = \sigma_1\).
Panel 3: as panel 1. Shifting the base point of loop \(\gamma_1\) by \(\Delta k_x = 2\pi\) changes the braid by a non-Abelian conjugate action, resulting in \(B_{\gamma_1^x} = B_x^{-1} B_{\gamma_1 } B_x = \sigma_1^{-1}\sigma_2^{-1}\sigma_1 = \sigma_2\sigma_1^{-1}\sigma_2^{-1}\).
Panel 4: as panel 1.
Subsequently shifting the base point by \(\Delta k_y = 2\pi\) leads to \(B_{\gamma_1}^{xy}=B_y^{-1} B_{\gamma_1}^x B_y = \sigma_1^{-1}\).
(c) Perturbation along \(H_s\) moves the EP2s across the boundary, fusing them after they have followed a non-trivial loop around the torus.
}
\label{fig:3-fusion}
\end{figure*}

{\it Non-Abelian fusion.---}
Fusion is the process of combining different point defects~\cite{RevModPhys.80.1083}. In Abelian systems, this leads to a unique outcome: for example, fusing two Weyl cones with opposite chiralities results in a trivial charge and the opening of a band gap~\cite{RevModPhys.90.015001}.
In contrast, non-Abelian fusion is path-dependent: the outcome depends on how the defects are brought together.

In our work, the unpaired EP3 is obtained by fusing two EP2s with opposite-signed winding numbers [see Fig.~\ref{fig:3-fusion}~(a)]. Furthermore, the fusion of two EP2s is path-dependent (or non-Abelian). Depending on the path taken to bring them together, two EP2s can yield either a topological braid, where an unpaired EP3 emerges, or a trivial braid, where a band gap opens. This path dependence serves as a clear signature of a non-Abelian topological phase transition triggered by moving EP2s along such nontrivial paths. While topological winding numbers fail to capture this transition, we successfully characterize it through braids. This is similar to bringing two Majoranas together, leading to either one fermion or zero fermions~\cite{RevModPhys.80.1083}.

To observe the non-Abelian fusion, we start by characterizing the braids of the two EP2s at \(\delta = 1\)~\cite{suppl}.
We show in Fig.~\ref{fig:3-fusion} that, relative to \(\delta=1+2\sqrt{2}\), the braids along the BZ boundary loop remain unchanged, but that instead of a single EP3 the system hosts two EP2s in the BZ, whose braid charges add up to the original charge of the upaired EP3.

We highlight the non-Abelian nature of the braid group by measuring the braid invariant for the same EP2 along different enclosing loops.
Shifting the base point of a loop \(\gamma\) that encircles an EP by another loop \(\phi\) is associated with a conjugate action \(B_\gamma \to B_\phi^{-1} B_\gamma B_\phi\)~\cite{bott1982differential} on the charge of the EP.
We experimentally confirm this dependence for a loop encircling one of the EP2s in Fig.~\ref{fig:3-fusion}~(b).
Taking \(k_x=k_y=-\pi\) as the base point, this EP2 corresponds to the braid \(B_{\gamma_1}=\sigma_2^{-1}\).
Shifting the base point along the \(k_x\)-direction to \(k_x=+\pi\), we find the braid \(B_{\gamma_1^x}=\sigma_2\sigma_1^{-1}\sigma_2^{-1}\), subsequently shifting it along the \(k_y\)-direction to \(k_x=k_y=+\pi\), we find \(B_{\gamma_1^{xy}}=\sigma_1^{-1}\).
The difference between these loops is a shift by a meridian or longitude around the torus.

Correspondingly, the movement of EP2s encircling these torus handles leads to different fusion results, highlighting their non-Abelian structure.
Next we study the process of bringing these two simple EP2s together along a different path in the BZ.
After splitting the unpaired EP3 into two EP2s, ending at \(H_M^{\delta = 1}\), with the two EP2s at $(k_x,k_y)\approx \pm(0.815,-0.815)$, we move these points around the BZ via the deformation
\(
    H_s := (1-s) H_M^{\delta=1}(k_x,k_y) + s H_M^{\delta=1}(\pi, k_y).
\)
We show this progression in Fig.~\ref{fig:3-fusion}~(c).
For increasing \(s\), the two EP2s move away from each other towards \((0,\pm\pi)\). They trace out a path cutting through the BZ and annihilate for \(s=3/4\). The overall difference between this fusion to a trivial degenerate point and the previous fusion into a braid-protected unpaired EP3 (in \(H_M^{\delta}\)) is one full encircling of the torus along the \(k_x\)-direction.

\begin{figure*}[ht!]
  \centering
\includegraphics[width=0.95\textwidth]{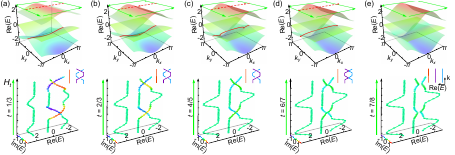}
\caption{
Topological transition between different band-gapped phases via creation and annihilation of a pair of EP2s.
eigenenergies of $H_t$ for (a) $t=1/3$ [the \((1,\sigma_2)\)-phase]), (b) $t=2/3$ (EP2 pair creation), (c) $t=4/5$ (EP2 encircling the BZ), (d) $t=6/7$ (EP2 pair annihilation), (e) $t=7/8$ [the \((1,1)\)-phase], respectively.
Continuous lines and surfaces correspond to theoretical predictions, while measured results are represented by dots;
error bars indicate the statistical uncertainty, obtained by assuming Poisson statistics in the photon-number fluctuations.
Top: Real parts of eigenenergies over the BZ, with Fermi arcs (solid lines) and their projections (dashed).
Bottom: Eigenenergy braids along the BZ boundary, counterclockwise from \((-\pi,-\pi)\).
}
\label{fig:4-transition}
\end{figure*}

By crossing the BZ boundary \(k_y=\pm\pi\) with increasing \(s\), the EP2s change the corresponding boundary braid from \(B_x=\sigma_1\) to the trivial braid \(B_x=1\).
This, in turn, changes the combined boundary braid enclosing the BZ to \( B_x B_y B_x^{-1} B_y^{-1} = 1 \sigma_2 1 \sigma_2^{-1}= 1\).
A trivial boundary braid allows for a gapped system, in accordance with our observation that the only two EP2s annihilate and vanish for \(s=1\) in Fig.~\ref{fig:3-fusion}(c).
Hence, after encircling the torus once, the EP2s fuse into a trivial degenerate point that can be gapped out perturbatively. These results confirm the path-dependent fusion rules.

The above process also suggests that the pair creation and fusion of EP2s, interspersed with topologically non-trivial paths through the BZ, correspond to non-Hermitian non-Abelian phase transitions~\cite{konigExceptionalTopologyNonorientable2025,WWY25}.
This is analogous to the conversion between topologically degenerate ground states via anyons~\cite{RevModPhys.80.1083}. Here non-Hermitian topological phases are defined and labeled based on the braiding of eigenenergies.
This arises in the so-called separation gap scheme, which is distinct from the usual K-theoretic phase classification in non-Hermitian band theory~\cite{KSU19,yangHomotopySymmetryNonHermitian2023}.
The non-Abelian topological phases discussed in our work are specified by two braids measured on loops following the meridian and longitude of the BZ torus or, more precisely, the conjugacy classes of such a pair of braids~\cite{PhysRevLett.120.146402,wojcikHomotopyCharacterizationNonHermitian2020,liHomotopicalCharacterizationNonHermitian2021,yangHomotopySymmetryNonHermitian2023,konigExceptionalTopologyNonorientable2025}.
The phase of \(H_{s=1}\) is, for example, denoted as \((B_x,B_y) = (1, \sigma_2)\), which contains an extensive Fermi arc [see Fig.~\ref{fig:4-transition}~(a)].

Taking \(H_{s=1}\) as a starting point, a phase transition to the completely trivial \((1,1)\)-phase is observed by measuring the spectrum of the Hamiltonian
\(
    H_{t}=(1-t) H_{s=1}(k_y)+t (2+\cos(k_x))\operatorname{diag}(1, 0, -1)
\)
for \(0\leq t\leq 1\)~\cite{suppl}.
As shown in Fig.~\ref{fig:4-transition}, both the initial and final phases are band-gapped, with the total boundary braid \(1 B_y 1 B_y^{-1} = 1\) being trivial.
However, the two phases cannot be brought into each other immediately, since the non-trivial \(B_y\) and corresponding extensive Fermi arc persist under perturbation.
Instead, a gapless topological transition that changes \(B_y\) must occur.
This transition proceeds via the pair creation of two EP2s at the origin \((k_x,k_y)=(0,0)\), which then encircle the BZ towards \((\pm\pi,0)\), respectively.
As \(t\) increases, the two EP2s gradually destroy the Fermi arc and eventually annihilate at \((\pm\pi,0)\), resulting in the trivial \((1,1)\)-phase without any braids or Fermi arcs at all.

{\it Discussion.---}
We have provided the first experimental realisation of a non-Abelian unpaired EP3. This novel degeneracy exploits a loophole in celebrated doubling theorems enabled by the highly intricate non-Abelian topology of non-Hermitian multi-band systems. We have further elucidated this topology through experimentally probing the fusion operations enabling the non-Abelian unpaired EP3. The experiment is facilitated by our design of new single-photon interferometric protocols, which enable the extraction of eigenstates and spectra of highly tunable non-Hermitian Hamiltonians.
Our setup opens up the possibility of simulating a wide class of topological phenomena in dissipative multi-band systems and probing fundamental aspects of non-Abelian topology that have thus far only been subject to theoretical inquiry.
The implications of these robust non-Abelian phenomena on dynamics, transport, and boundary effects are of great interest for future study. For instance, in the context of the chiral state transfer near EPs~\cite{ZWH18,CAJ21,NLL22,RLZ22}, the non-Abelian fusion demonstrated here enables the path-dependent creation and annihilation of higher-order EPs, offering a new form of control for multi-mode switching.

{\it Note added.---}Recently, a related work appeared~\cite{WWY25}.

\begin{acknowledgments}{\it Acknowledgments.---}
The authors thank Jan Carl Budich and Zhi Li for stimulating discussions.
This work has been supported by the National Key R\&D Program of China (Grant No. 2023YFA1406701) and National Natural Science Foundation of China (Grant Nos. 12025401, 92265209, 12374479, 12474352, 92476106).
KY is supported by the ANR-DFG project (TWISTGRAPH).
JLKK and EJB were supported by
the Swedish Research Council (VR, grant 2018-00313),
the Wallenberg Academy Fellows program (2018.0460),
and the project Dynamic Quantum Matter (2019.0068) of the Knut and Alice Wallenberg Foundation,
as well as the G\"{o}ran Gustafsson Foundation for Research in Natural Sciences and Medicine.
KKW acknowledges support from the Natural Science Foundation of Anhui Province (2508085Y002).
KKW and LX acknowledge support from Beijing National Laboratory for Condensed Matter Physics (No. 2024BNLCMPKF010).
\end{acknowledgments}

\bibliography{your-references-here}

\onecolumngrid
\vspace{1em}
\noindent
{\centering{\bf End Matter}\par}
\vspace{1em}
\twocolumngrid

\setcounter{equation}{0}
\renewcommand{\theequation}{B\arabic{equation}}

{\it Initial state preparation.---}
We prepare the initial state \(\rho_i\) as the completely mixed state $\rho_i=\one/3$, where $\one$ is the $3\times3$ identity matrix.
Starting from horizontally polarised light, we first tune the polarisation of the single photons through a half-wave plate (HWP) with a setting angle of $\cos^{-1}(\sqrt{2/3})/2 \approx 17.6^\circ$, creating \(\sqrt{2/3}\ket{H}+\sqrt{1/3}\ket{V}\).
We then destroy the coherence between the polarisation states of the photons using an unbalanced interferometer (UBI) composed of two polarizing beam splitters (PBSs) and two mirrors~\cite{unbalancedMZI}, leading to state \(\sqrt{2/3}\ketbrad{H}+\sqrt{1/3}\ketbrad{V}\).
Subsequently, we separate the beam into two spatial modes, using a beam displacer (BD), which transmits vertically polarized photons into the upper spatial mode while horizontally polarized photons undergo a \(3\,\)mm lateral displacement into the lower mode.
Using another two HWPs (with setting angles of $45^\circ$ and $22.5^\circ$)  and a second UBI, we destroy the coherence in the lower spatial mode.
This results in the completely mixed photon state $\rho_i$ in the space spanned by the horizontally polarised upper mode \(\ket{H_U}\), the horizontally polarized lower mode \(\ket{H_L}\), and the vertically polarised lower spatial mode \(\ket{V_L}\).

{\it Realisation of non-unitary evolution.---}
Both eigenstate preparation \(U_1\) and non-Hermitian evolution \(U_2\) are non-unitary operations.
Implementing such arbitrary non-unitary evolution experimentally is challenging due to the difficulties in achieving gain in quantum systems~\cite{MaguancongNP2022,StefanoEPL17}. We bypass this obstacle by mapping to passive evolution without gain.
For a given evolution \(\tilde{U}(\tau)\) with gain, we map to \(U=\tilde{U}(\tau)e^{-\Lambda(\tau)}\).
We choose \(\Lambda(\tau)=\ln\sqrt{\max_j{\abs{\lambda_j}}}\), where \(\lambda_j\) are the eigenenergies of \(\tilde{U}(\tau) \tilde{U}^\dagger(\tau)\).
Under this mapping, gain terms are converted to no loss, and loss terms to even greater loss, leading to a passive operation \(U\)~\cite{Xue21,Xue23,YHL23}.

We implement such passive non-unitary evolution \(U\) using its singular-value decomposition $U=VDW$~\cite{TRS18}.
Here, the operator $D$ is a non-unitary diagonal matrix with the first diagonal element $D_{11}=1$, and the rest less than $1$ (see Fig.~1 in the main text).
It is realized by introducing mode-selective losses of photons by adjusting the angles of $\mathrm{H}_1$ and $\mathrm{H}_2$ shown in Fig.~1 in the main text.
The unitary operations of $W$ and $V$ can be further decomposed as $W=W_{12}W_{13}W_{23}$ and $V=V_{23}V_{13}V_{12}$, such that the unitary operators $W_{ij}$ and $V_{ij}$ only act on a two-dimensional subspace of the qutrit system~\cite{RZB94}.
We realise them by combining photons from different hybrid modes into a certain spatial mode, and then applying the transformation through wave plates.

{\it Eigenstate Extraction.---}
We purify the initial mixed state $\rho_i$ into a right eigenstate \(\ket{R_j}\) of $H(\textbf{k})$ using an appropriately chosen non-unitary evolution \(U_1 = e^{f_j(H)\tau_1}\).
As a concrete example, we prepare the eigenstate \(\ket{R_1}\) with the greatest real part of the energy, using the imaginary-time evolution governed by $f_1(H)=H$.
The corresponding evolution operator is
\begin{equation}
\label{eq:psi1}
    U_1 = e^{H\tau_1}
    = \sum_j e^{E_{j}\tau_1} \Pi_j ,
\end{equation}
where $E_{j}$ are the eigenenergies of $H$, and \(\Pi_j\) are the projectors to the corresponding eigenspace, satisfying \(\Pi_j\Pi_k = \delta_{jk}\Pi_k\) and \(\sum_j \Pi_j = \one\).
Away from the EPs, the projectors can be expressed as \(\Pi_j = \ket{R_j}\bra{L_j}/\braket{L_j}{R_j}\) in terms of the corresponding right (left) eigenstates $\ket{R_j}$ ($\bra{L_j}$) of \(H\).
These eigenstates satisfy $\braket{L_j}{R_k}_{j\neq k}=0$, and we normalise them as $\braket{L_j}{L_j}=\braket{R_j}{R_j}=1$.
We write the real and imaginary parts of the eigenenergies as $E_{j}=E_{j}^{R}+iE_{j}^{I}$, and sort them such that for $j<k$, $E_{j}^{R}>E_{k}^{R}$ without loss of generality.

After evolving under $U_1$, the initial state $\rho_i$ becomes
\begin{align}
    \rho_{\tau_1}
    =U_1(\tau_1)\rho_iU^{\dagger}_1(\tau_1)
    =\sum_{j,k} e^{\tau_1(E_{j}+E_{k}^*)}\frac{c_{jk}}{3}\ketbra{R_j}{R_k},
\label{eq:rhot}
\end{align}
where $c_{jk}=\braket{L_j}{L_k}/(\braket{L_j}{R_j}\braket{R_k}{L_k})$.
The exponential \(e^{\tau_1(E_{j}+E_{k}^*)}=e^{(E_{j}^R+E_{k}^R)\tau_1}e^{i(E_{j}^I-E_{k}^I)\tau_1}\) damps all eigenstates relative to \(\ket{R_1}\), which corresponds to the eigenenergy with the largest real part, and which is therefore the steady state.

Likewise, by using $f_3(H)=-H$ in $U_1$, we obtain as the steady state $\ket{R_3}$, which has the lowest real eigenenergy of $E_{3}^R$.
To determine the eigenstate $\ket{R_2}$, we first obtain the eigenenergies $E_{1}$ and $E_{3}$ by interferometric measurements. We formulate a new Hamiltonian $\tilde H = [H-(E_{1}+E_{3})\one/2 ]^2$, which can arise from nonlinearity in the physical implication~\cite{PSM24}.
While this retains the same eigenstates, the real components of the corresponding eigenenergies of $\ket{R_1}$ and $\ket{R_3}$ both become $[(E_{3}^R-E_{1}^R)^2-(E_{3}^I-E_{1}^I)^2]/4$.
Consequently, we can obtain $\ket{R_2}$ as the steady pure state of the imaginary time evolution governed by $f_2(H) = -\tilde H$ or $f_2(H) = \tilde H$.
To purify the evolved states with high fidelity, we fix the evolution time $\tau_1=10$ for the eigenstate extraction.

{\it Eigenenergy Measurement.---}
The single photons, prepared in the eigenstate $\ket{R_j}$, are injected into a non-polarizing beam splitter (NPBS). It splits the beam into two spatial modes, transmitted $\ket t$ and reflected $\ket r$.
We apply the non-unitary operation $U_2=e^{-i(H-i\Lambda\one)\tau_2}$ on the photons in the transmitted mode $\ket t$, while leaving those in $\ket r$ unchanged. The state then evolves to $(e^{-i\tilde E_{j}\tau_2}\ket{t}\ket{R_j}+\ket{r}\ket{R_j})/\sqrt{2}$, where $\tilde E_{j}= E_{j}-i\Lambda$. Without loss of generality, we fix $\tau_2=1$ for $U_2$.

Subsequently, we perform interferometric measurements to obtain the eigenenergies $E_{j}$ of $H(\textbf{k})$~\cite{Xue23}.
We achieve this by first applying a HWP at 45$^\circ$ to the transmitted photons. Then, the transmitted and reflected photons are recombined using a PBS.
Thereby, a generic state $\ket{R_j}=\alpha_j\ket{H_U}+\beta_j\ket{H_L}+\gamma_j\ket{V_L}$ evolves into
\begin{align}
\ket{R_j'}=&\frac{1}{\sqrt{2}}\left(\alpha_j\ket{t' U}+\beta_j\ket{t'L}\right)\left(\ket{H}+e^{-i\tilde E_{j}}\ket{V}\right)\\ \nonumber
&+\frac{1}{\sqrt{2}}\gamma_j\ket{r'L}\left(e^{-i\tilde E_{j}}\ket{H}+\ket{V}\right).
\end{align}
Here, $\ket{t'}$ and \(\ket{r'}\) denote the transmitted (reflected) modes after passing the photons reflected by the NPBS through the second PBS.

Finally, we measure the complex phase through successive projective measurements on the polarisation states $\{\ket{\pm}=\left(\ket{H}\pm\ket{V}\right)/\sqrt{2}, \ket{\circlearrowleft}=\left(\ket{H}-i\ket{V}\right)/\sqrt{2}\}$ in the different spatial modes.
Measuring, e.g., coincidence counts \(N^+_{t'U}\) for the state \(\ket{+,t'U}\), normalised to the total number \(N_\text{tot}\) of input photons after purification, leads to
\begin{align}
    \frac{N^+_{t'U}}{N_{\text{tot}} }
    &= \abs{\bra{t'U,+} \ket{R'_j}}^2
    \\&= \notag
    \abs{\frac{\alpha_j}{2}(1+e^{-i \tilde E_j})}^2
    \\&= \notag
    \frac{\abs{\alpha_j}^2}{2}
        \left(\frac{1+e^{2\Im (\tilde E_j)}}{2}
        + \frac{e^{-i \tilde E_j}+e^{i \tilde E_j^*}}{2} \right).
\end{align}
The linear combination
\begin{align}
\xi_j=
\Bigg[&
\left(
    (i+1)\frac{N_{t'U}^+}{N_\text{tot}}
    +(i-1) \frac{N_{t'U}^-}{N_\text{tot}}
    -2i \frac{N^\circlearrowleft_{{t'U}}}{N_\text{tot}}
    \right)
\\ \nonumber
&+
\left(
    (i+1) \frac{N_{t'L}^+}{N_\text{tot}}
    +(i-1) \frac{N_{t'L}^- }{N_\text{tot}}
    -2i \frac{N^\circlearrowleft_{t'L}}{N_\text{tot}}
    \right)
\\ \nonumber
&+
\left(
    (1-i) \frac{N_{r'L}^+ }{N_\text{tot}}
    -(i+1) \frac{N_{r'L}^- }{N_\text{tot}}
    +2i \frac{N^\circlearrowleft_{r'L}}{N_\text{tot}}
    \right)
\Bigg]
\end{align}
of such coincidence counts equals \(\xi_j=\expval{U_2}{R_j'}\), from which we finally obtain the eigenenergies of $H{(\textbf{k})}$ with
\begin{equation}
    E_j=\tilde E_j+i\Lambda=i\ln(\xi_j)+i\Lambda.
\end{equation}

{\it Definition of the winding number.---}
The winding number $W_\mathcal{C}$ can be obtained by counting the number of strand exchanges, taking their direction into account but disregarding which bands are involved.
This reduction corresponds to the Abelianisation of the three-strand braid group $\mathbb B_3/[\mathbb B_3,\mathbb B_3]= \mathbb Z$ \cite{konigBraidprotectedTopologicalBand2023,yangHomotopySymmetryNonHermitian2023}.
We can equivalently express it in terms of the discriminant of the characteristic polynomial \cite{yangFermionDoublingTheorems2021,patilMeasuringKnotNonHermitian2022,yangHomotopySymmetryNonHermitian2023},
\begin{align}
    \frac{\mathbb B_3}{[\mathbb B_3,\mathbb B_3]}:\, W_{\mathcal C}&=\sum_{i\ne j}\frac{1}{2\pi}\oint_{\mathcal C} \partial_{\mathbf k}\operatorname{arg}\left[E_i(\mathbf k)-E_j(\mathbf k)\right]d\mathbf k\nonumber\\
    \label{eq:discriminant}
    &=\frac{1}{2\pi i}\oint_{\mathcal C}\partial_{\mathbf k}\ln \operatorname{disc}[H(\mathbf k)]d\mathbf k.
\end{align}
The discriminant $\operatorname{disc}[H(\mathbf k)]=\prod_{i<j}[E_i(\mathbf k)-E_j(\mathbf k)]^2$ is a non-zero complex number for non-degenerate matrices, and the braid degree is its winding number from the fundamental group of the punctured complex plane $W_\mathcal{C}\in\pi_1(\mathbb C-\{0\})=\mathbb Z$.

\clearpage
\begin{widetext}
\appendix

\renewcommand{\thesection}{\Alph{section}}
\renewcommand{\thefigure}{S\arabic{figure}}
\renewcommand{\thetable}{S\Roman{table}}
\setcounter{figure}{0}
\renewcommand{\theequation}{S\arabic{equation}}
\setcounter{equation}{0}

\section{Supplemental Material for ``Observation of Braid-Protected Unpaired Exceptional Points"}
\maketitle

In the Supplemental Material, we provide additional details and data supporting the main text. In addition, we propose an extension of our setup that enables the measurement of the eigenstates and spectra of arbitrary $N$-level non-Hermitian Hamiltonians.

\subsection{Robustness of the braids.}
\begin{figure*}[b!]
\centering
\includegraphics[width=0.4\textwidth]{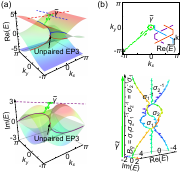}
\caption{
Observation of the unpaired third-order exceptional point (EP3) braid along an irregular loop $\tilde{\gamma}$. (a) Theoretical results for the real and imaginary parts of the spectrum of $H_M^{\delta=1+2\sqrt{2}}$. The red dot marks the non-Abelian unpaired EP3. (b) Top: Parametric irregular loop $\tilde{\gamma}$ based at ($k_x$, $k_y$) = ($-\pi, -\pi$). Bottom: Eigenvalue braid of $H_M^{\delta=1+2\sqrt{2}}$ along this loop. Theoretical predictions are shown as solid lines, while experimental data are plotted as points. Error bars indicate statistical uncertainties, assuming Poisson statistics for photon-number fluctuations.
}
\label{fig:S1-robustness}
\end{figure*}

In our work, the topological robustness of the braids is demonstrated by deforming the measurement loop in the Brillouin zone (BZ). Specifically, the braid $B_{\gamma}$ obtained in our experiment remains invariant when the loop is deformed to the boundary of the BZ, as shown in Fig.~2(c–d) of the main text.

To further emphasize this robustness, we perform an additional experiment along an irregular loop $\tilde{\gamma}$ encircling the unpaired third-order exceptional point (EP3). This loop traverses counterclockwise along the line $k_y = k_x + R$, with a circular segment of radius $0.1\pi + R$, where $R \in [-0.1,0.1]$ is randomly sampled for each point along the line. Along this irregular, perturbed path, the same braid pattern is observed (see Fig.~\ref{fig:S1-robustness}), further confirming the topological robustness of the braids.

\subsection{Splitting of the unpaired exceptional points.}

\begin{figure*}
  \centering
\includegraphics[width=0.95\textwidth]{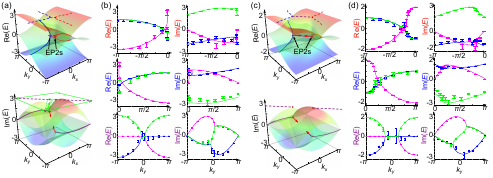}
\caption{Observation of Fermi and i-Fermi arcs in $H_{\delta}^M$.
Theoretical results for the real and imaginary parts of the eigenenergies of $H_M^{\delta}$ with (a) $\delta=1.5$ and (c) $\delta=1$ are shown as colored surfaces over the Brillouin zone (BZ). The red dots correspond to the twofold exceptional points (EP2s). The solid lines depict both the Fermi arcs of the real eigenenergy components and the i-Fermi arcs of the imaginary components. The dashed lines represent their projections on the $k_x-k_y$ plane. Measured complex eigenenergies along the Fermi and i-Fermi arcs drawn in (a) and (c) are presented in (b) and (d), respectively. Experimental data are represented by symbols, while theoretical predictions are denoted by colored lines. Error bars indicate the statistical uncertainty, obtained by assuming Poisson statistics in the photon-number fluctuations.}
\label{fig:suppl1}
\end{figure*}

\begin{figure*}
  \centering
\includegraphics[width=0.95\textwidth]{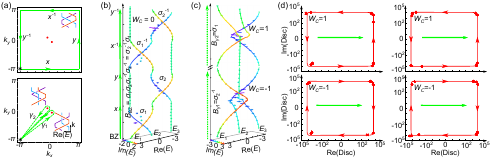}
\caption{The braids in $H_{\delta}^M$ with $\delta=1.5$.
(a) Parametric loops along the BZ boundary (green solid lines in the top panel) and EP2s of $\gamma_{1,2}$ (green dashed lines in the bottom panel). Exceptional points are represented by red dots. The corresponding measured braids are shown in (b) and (c), respectively. (d) Measured discriminant for the four segments along the boundary of the BZ. Experimental data are presented as symbols, while theoretical predictions are denoted by colored lines. Error bars indicate statistical uncertainty, obtained by assuming Poisson statistics in the photon-number fluctuations.}
\label{fig:suppl2}
\end{figure*}

\begin{figure*}
  \centering
\includegraphics[width=0.66\textwidth]{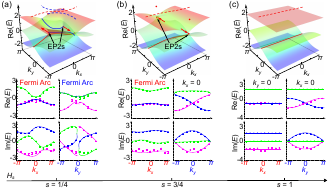}
\caption{Observation of the eigenenergies of $H_{s}$.
The measured real and imaginary parts of the bands of $H_{s}$ with $s=1/4$ (a), $s=3/4$ (b), and $s=1$ (c). Top: Theoretical results for the real and imaginary parts of the eigenenergies are shown as colored surfaces over the BZ. The red dots correspond to EP2s. The solid lines depict the Fermi arcs, while the dashed lines are their projections onto the $k_x-k_y$ plane. Bottom: Measured complex eigenenergies along the Fermi arcs or $k_x=0$ and $k_y=0$. Experimental data are presented as symbols, while theoretical predictions are denoted by colored lines. Error bars indicate statistical uncertainty, obtained by assuming Poisson statistics in the photon-number fluctuations.}
\label{fig:suppl3}
\end{figure*}

When the parameter $\delta$ in $H_M^{\delta}$ is tuned to be less than $1+2\sqrt{2}$, the unpaired EP3 splits into a pair of second-order exceptional points (EP2s). Figure~\ref{fig:suppl1} illustrates that as $\delta$ decreases to $1.5$ and $1$, these EP2s can be found at $(k_x, k_y)\approx\pm(0.231, -0.231)$ and $(k_x, k_y)\approx\pm(0.815, -0.815)$, respectively.  It is noticeable that as $\delta$ decreases, the pair of EP2s moves away from each other along the line of $k_x=-k_y$.
Moreover, with the decrease of $\delta$, the two Fermi arcs (signifying the closure of one of the real energy gaps $\text{Re}(E_i-E_j)=0$) intersecting at the unpaired EP3 split to cross each EP2. The i-Fermi arc (representing the closure of one of the imaginary energy gaps) along $k_x=-k_y$ is separated by a new emerging Fermi arc that terminates at the EP2s. In Figs.~\ref{fig:suppl1}(b) and (d), we show the measured real and imaginary parts of the bands along these Fermi and i-Fermi arcs, illustrating the splitting of the unpaired EP3.

When encircling an exceptional point along a counterclockwise path, the eigenenergies start to evolve around each other. This pattern forms a braid on the eigenenergies. If the path can be constructed by consecutive loops, the total braid pattern is the product of the braids around each loop. This establishes a natural correspondence between exceptional points and braid patterns, where each exceptional point can be associated with an element in the braid group.
For a general $N$-band non-Hermitian system, the eigenenergies can be sorted as $\text{Re}(E_j)\geq\text{Re}(E_{j+1})$. The braid group is constructed by the braids $\sigma_j$ ($\sigma_j^{-1}$) ($1\leq j \leq N-1$), which swap $E_j$ and $E_{j+1}$ counterclockwise (clockwise) in the complex plane. They satisfy the Yang-Baxter relation and far commutativity
\begin{equation}
\sigma_j\sigma_{j+1}\sigma_j=\sigma_{j+1}\sigma_{j}\sigma_{j+1},\ \sigma_{i}\sigma_{j}=\sigma_{j}\sigma_{i} \ (|i-j|>1).
\label{eq:s1}
\end{equation}
Similar to the Berry flux/Chern number of the eigenstates when wrapping around a Weyl cone, this braid invariant of the complex spectra when going around an EP is the topological charge associated with the EP.  It determines the consequences of putting EPs together or splitting them up.

As shown in Figs.~2(c) and (d) in the main text, we measure the braids of the unpaired EP3 along the loops of $\gamma$ and the boundary of the BZ. They can be represented by $\sigma_1\sigma_2\sigma_1^{-1}\sigma_2^{-1}$, where the deformations of the loop do not alter the braid. As also shown in Fig.~\ref{fig:suppl2} and Fig.~3(b), the splitting of the unpaired EP3 by tuning the on-site potential to $\delta=1.5$ and $\delta=1$ does not alter the braid along the boundary of the BZ.

Additionally, we measure the braid patterns when encircling each of the EP2s, observing that the braids are topologically invariant as $\delta$ decreases. Specifically, the EP2s at $(k_x\approx0.231, k_y\approx-0.231)$ for $\delta=1.5$ and $(k_x\approx0.815, k_y\approx-0.815)$ for $\delta=1$ braid the second two eigenenergies clockwise along path $\gamma_1$. In contrast, the EP2s located at $(k_x\approx-0.231, k_y\approx0.231)$ and $(k_x\approx-0.815, k_y\approx0.815)$ braid the first two eigenenergies counterclockwise along $\gamma_2$. Consequently, the braids of EP2s exhibit the non-Abelian braids of $\sigma_2^{-1}$ and $\sigma_1$,
compensating the boundary braids according to the non-Abelian sum rule. The corresponding winding numbers of these braids are $\mp1$ for $\gamma_{1,2}$ and 0 along the boundary of the BZ.

Our experiment verifies the Yang-Baxter relation directly for three bands,
as the EP2s with charges \(\sigma_2^{-1}\) and \(\sigma_1\) fuse to EP3 with charge \(\sigma_1\sigma_2\sigma_1^{-1}\sigma_2^{-1}\). Equality between these, \(\sigma_2^{-1}\sigma_1=\sigma_1\sigma_2\sigma_1^{-1}\sigma_2^{-1}\) implies immediately that \(\sigma_1\sigma_2\sigma_1=\sigma_2\sigma_1\sigma_2\).

If one imagines the three-band system in our experiment as a subsystem of a larger Hilbert space in which all other bands are trivial, it may also be intuitively clear that the Yang-Baxter relation holds for all other generators \(\sigma_i, \sigma_{i+1}\).

Far commutativity becomes relevant for the braid group \(B_{n\geq 4}\), i.e., for systems of four or more bands. A simple model showcasing this, we have
\begin{equation}
    H_\text{far-comm.}=\left(\begin{array}{cccc}
    2&1 &0&0\\
    e^{i k_x} & 2 &0&0\\
    0&0 & -2&1\\
    0&0 & e^{i k_y} & -2
    \end{array}\right),
\end{equation}
with boundary braids \(B_x = \sigma_1, B_y = \sigma_3\). It is gapped everywhere, which implies trivial \(B_\text{BZ} = \sigma_1\sigma_3\sigma_1^{-1}\sigma_3^{-1}\), and proves far commutativity.


Figure~\ref{fig:suppl3} illustrates the measured Fermi arcs in $H_s$ as $s$ increases. The initial model $H_{s=0}=H_{\delta=1}^{M}$ [shown in Figs.~\ref{fig:suppl1}(c) and (d)] hosts two topologically nontrivial Fermi arcs, each along a great circle. With increasing $s$, these two Fermi arcs are cleaved into three segments by the exceptional points. Upon the annihilation of the EP2s with $s>3/4$, two of the Fermi arcs, terminated by exceptional points, are eliminated from the Brillouin zone.

\subsection{Topological transition between different band-gapped phases.}

\begin{figure*}
  \centering
\includegraphics[width=0.95\textwidth]{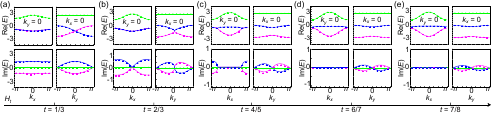}
\caption{
Observation of the eigenenergies of $H_{t}$. Measured real and imaginary parts of the eigenenergies of $H_{t}$ at $k_y=0$ and $k_x=0$ with different $t$ are shown in (a-e). Continuous lines correspond to theoretical predictions, while measured results are represented by dots. Error bars indicate statistical uncertainty, obtained by assuming Poisson statistics in the photon-number fluctuations.}
\label{fig:suppl4}
\end{figure*}

As shown in Fig.~\ref{fig:suppl4}, we measure the spectrum of the Hamiltonian
$H_{t}=(1-t) H_{s=1}(k_y)+t \,(2+\cos k_x)\,\text{diag}(1,0,-1)$ for $k_y=0$ and $k_x=0$ at different fixed values of $t$. At $t=1/3$, the spectrum exhibits a Fermi arc along $k_y=0$ [Fig.~\ref{fig:suppl4}(a)]. As $t$ increases, a phase transition to a fully trivial phase occurs, marked by the disappearance of the Fermi arc. This transition proceeds via the pair creation of two EP2s at the origin $(k_x,k_y)=(0,0)$ [Fig.~\ref{fig:suppl4}(b)], which then move along paths encircling the BZ towards $(\pm\pi,0)$ [Figs.~\ref{fig:suppl4}(c–d)]. The two EP2s gradually destroy the Fermi arc and eventually annihilate at $(\pm\pi,0)$ [Fig.~\ref{fig:suppl4}(e)].

\subsection{Experimental implementation.}
In our experiment, a three-level system is encoded in the hybrid space of the polarization and two spatial modes of single photons generated via type-II spontaneous parametric down-conversion. In this framework, the basis states are encoded as $\{\ket{H_U}=(1,0,0)^\text{T}, \ket{H_L}=(0,1,0)^\text{T}, \ket{V_L}=(0,0,1)^\text{T}\}$, where $H/V$ denote the horizontal/vertical polarizations and $U/L$ denote the upper/lower spatial modes. After preparing the initial state as the completely mixed state $\rho_i=1/3(\ketbrad{H_U}+\ketbrad{H_L}+\ketbrad{V_L})$, we implement the non-unitary operator $U_1$ to purify $\rho_i$ into one of the right eigenstates $\ket{R_j}$ of $H(\textbf{k})$.

Here we give a concrete example of the experimental process. For the Hamiltonian $H^{\delta=1+2\sqrt{2}}_M$ with quasimomenta $(k_x=-0.49\pi, k_y=-0.65\pi)$, we obtain
\begin{align}
H^{\delta=1+2\sqrt{2}}_M=\begin{pmatrix}
3.7970 + 0.9995i & -0.9995 - 1.0314i & 0\\
-0.9995 - 1.0314i & 0.4854 - 0.1085i & -0.8910 - 0.5460i\\
0& -0.8910 - 0.5460i & -4.2824 - 0.8910i
\end{pmatrix}.
\end{align}
In our experiment, we apply the non-unitary operator $U_1=e^{\tau H}$ with evolution time $\tau=10$ to obtain the eigenstate $\ket{R_1}$. To circumvent the difficulty of implementing gain, we map $U_1$ to
\begin{align}
U'_1=U_1e^{-\Lambda(\tau=10)}=\begin{pmatrix}
-0.6554 + 0.5733i & 0.3145 - 0.1097i & -0.0402 + 0.0027i\\
0.3145 - 0.1097i & -0.1272 - 0.0061i & 0.0147 + 0.0048i\\
-0.0402 + 0.0027i& 0.0147 + 0.0048i & -0.0016 - 0.0010i
\end{pmatrix}
\end{align}
by converting gain terms to no loss and loss terms to greater loss. Here $U'_1$ can be decomposed as $U'_1=VDW$ using singular-value decomposition. The unitary operators $V$ and $W$ are
\begin{align}
W=\begin{pmatrix}
-0.7024 + 0.6144i & 0.3370 - 0.1175i & -0.0431 + 0.0029i\\
-0.1407 - 0.3287i & -0.0245 - 0.9098i & -0.0904 + 0.1887i\\
-0.0284 + 0.0237i& -0.2097 + 0.0165i & -0.9074 - 0.3619i
\end{pmatrix}
\end{align}
and
\begin{align}
V=\begin{pmatrix}
0.9332 & -0.3575 & -0.0378\\
-0.3310 - 0.1334i & -0.8456 - 0.3354i & -0.1746 - 0.1212i\\
0.0344 + 0.0262i & 0.1401 + 0.1585i & -0.4762 -0.8524i
\end{pmatrix}.
\end{align}
The operator $D=\mathrm{diag}(1,0,0)$.

The unitary operators are further decomposed as $W=W_{12}W_{13}W_{23}$ and $V=V_{23}V_{13}V_{12}$. The decomposed operators $W_{ij}$ are given by
\begin{align*}
W_{23}&=\begin{pmatrix}
1 & 0 & 0\\
0 & -0.6627 - 0.7187i & 0.0595 + 0.2019i\\
0 & 0.2099 - 0.0166i & 0.9080 + 0.3621i
\end{pmatrix},
W_{13}=\begin{pmatrix}
0.9993 & 0 & -0.0284 - 0.0237i\\
0 & 1 & 0\\
-0.0284 + 0.0237i & 0 & -0.9993
\end{pmatrix},\\
W_{12}&=\begin{pmatrix}
-0.7029 + 0.6148i & 0.1408 + 0.3289i & 0\\
-0.1408 - 0.3289i & 0.7029 + 0.6148i & 0\\
0 & 0 & 1
\end{pmatrix}.
\end{align*}
Each operator $W_{ij}$ only acts nontrivially on a two-dimensional subspace of the three-level system. As shown in Fig.~1 of the main text, $W_{23}$ is implemented by inserting a sandwich-type sequence of quarter-wave plates (QWPs) and half-wave plates (HWP), i.e., QWP-HWP-QWP (QHQ), in the lower spatial mode. To realize $W_{13}$, a beam displacer (BD) is used to combine the states $\ket{H_U}$ and $\ket{V_L}$ into the upper spatial mode, followed by a QHQ. Subsequently, a HWP oriented at 45$^\circ$ is inserted in the lower spatial mode, and together with a BD, combines $\ket{H_U}$ and $\ket{H_L}$ into the lower spatial mode. Finally, another QHQ is used to implement the operator $W_{12}$. To implement $D$, the angles of HWPs $H_1$ and $H_2$ are both set to 0$^\circ$. Following this, the operators $V_{ij}$ are similarly realized using combinations of wave plates and BDs.

After preparing the photons in the state of $\ket{R_1}$, we use a nonpolarizing beam splitter (NPBS) to split the beam into transmitted and reflected spatial modes. Subsequently, in a manner similar to the implementation of $U'_1$, the non-unitary operation $U_2$ is applied in the transmitted mode, while the photon state in the reflected mode remains unchanged. The corresponding eigenenergy	$E_1$ is then obtained via interferometric measurements.

Analogously, by adjusting $U_1$ accordingly, we obtain the eigenenergy $E_3$ through interferometric measurements. Using the measured eigenenergies $E_1$ and $E_3$, we construct a new Hamiltonian $\tilde{H}$. By performing imaginary-time evolution governed by $\tilde{H}$ to obtain the eigenstate $\ket{R_2}$, the remaining eigenenergy $E_2$ can also be determined.

In this work, we employ single photons in an interferometric platform to investigate braid-protected unpaired EPs. Leveraging the quantum nature of photons and precise interferometric control, our approach provides full access to both amplitude and phase information, enabling a complete characterization of the eigenstates of the non-Hermitian Hamiltonian as well as direct interferometric measurements of the corresponding eigenvalues. This capability enables an unambiguous identification of EPs and the direct observation of braiding and fusion processes. Moreover, the linear quantum-optical architecture offers full and independent tunability of all system parameters, realizing a highly reconfigurable simulator for non-Hermitian Hamiltonians and enabling systematic exploration of a broad class of non-Abelian braiding phenomena within a single experimental platform.

Moreover, while single-photon behavior may resemble that of classical waves in some cases, the latter only simulates the behavior of quantum mechanics. Performing the experiment with single photons allows the preparation and evolution of genuine quantum states with well-defined quantum coherence. These are essential for realizing and probing non-Abelian topological effects in the genuine quantum regime. They are also crucial for future studies and applications where the impact of quantum noise, many-body effects, and entanglement on non-Abelian topology is expected to warrant further attention.

\begin{figure*}
  \centering
\includegraphics[width=0.6\textwidth]{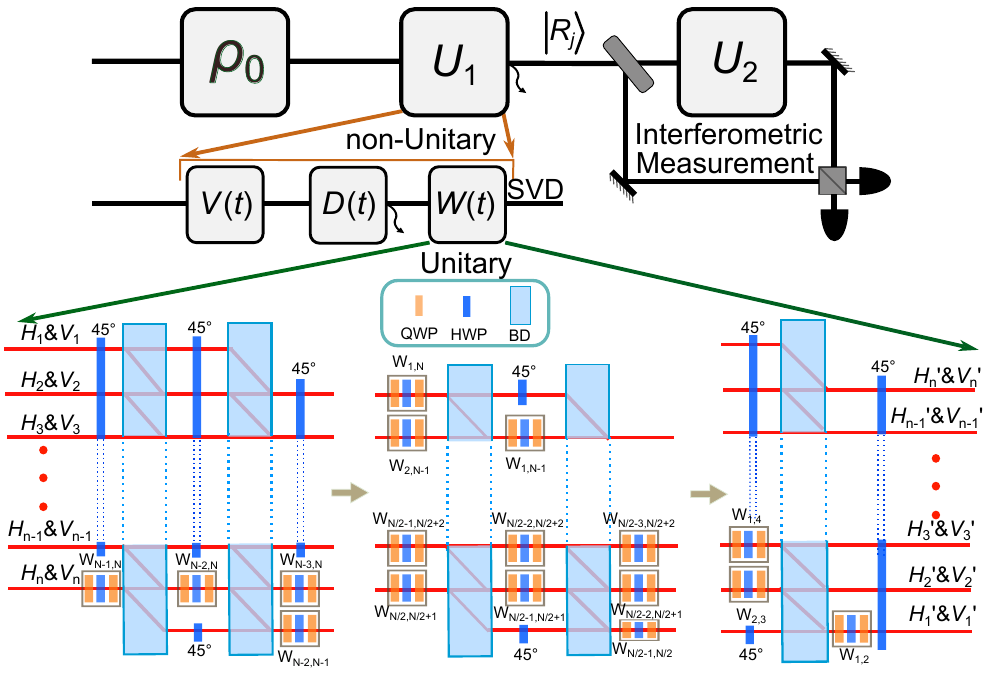}
\caption{
Experimental scheme for measuring eigenstates and eigenenergies of an arbitrary $N$-level non-Hermitian Hamiltonian. The system is first initialized in a mixed state $\rho_0$, which is then purified into an eigenstate $\ket{R_j}$ via a non-unitary operation $U_1$. To extract the corresponding eigenvalue, a second non-unitary operation $U_2$ governed by $H$ is applied in one of two interferometer paths, inducing a relative complex phase shift. Interferometric measurements read out the eigenvalue. Arbitrary non-unitary operations $U_1$ and $U_2$ can be decomposed via singular-value decomposition (SVD) into two unitary operations $W$ and $V$, combined with a diagonal matrix $D$, which can be realized using interferometer networks.
}
\label{fig:suppl5}
\end{figure*}

\subsection{Generalization to $N$-level Non-Hermitian Hamiltonian.}
Our setup can be naturally extended to observe the eigenstates and spectra of arbitrary $N$-level non-Hermitian Hamiltonians by exploiting the scalable photonic degrees of freedom and flexible interferometric architectures.

As illustrated in Fig.~\ref{fig:suppl5}, the procedure begins with preparing the initial maximally mixed state $\rho_0$. For an $N$-level system, basis states can be encoded into spatial and polarization modes of single photons. For even $N=2n$, the mapping reads
\begin{align*}
\left(\ket{0}, \ket{1}\right),\left(\ket{2}, \ket{3}\right),\cdots,\left(\ket{N-1}, \ket{N}\right)\Longleftrightarrow
\left(\ket{H_1}, \ket{V_1}\right),\left(\ket{H_2}, \ket{V_2}\right),\cdots,\left(\ket{H_n}, \ket{V_n}\right),\nonumber
\end{align*}
where $n=N/2$ denotes the number of the spatial modes and $H$ ($V$) denotes the horizontal (vertical) polarization of the single photons. For odd $N=2n-1$, one additional spatial mode encodes the unpaired state.

The preparation of such mixed states can be realized by successively splitting spatial modes with BDs and adjusting weights through HWPs. The coherence can be destroyed by using unbalanced interferometers. For an $N$-level system, only $\mathcal{O}(N)$ optical elements are required.

To obtain a specific eigenstate, we employ a non-unitary evolution $U_1 = e^{f_j(H)\tau_1}$ with adjustable evolution time $\tau_1$. Under appropriate filtering Hamiltonian $f_j(H)$, all but the $j$-th eigenstate decay, yielding the purified state $\ket{R_j}$. By scanning $j$, the complete set of eigenstates can be reconstructed. The corresponding complex eigenenergies are then obtained by applying a second non-unitary operator $U_2$ governed by $H$ in one arm of an interferometer. This evolution imprints the eigenvalue as a complex phase shift relative to the reference path, which is subsequently measured interferometrically.

General non-unitary evolutions can be implemented experimentally by decomposing them into a sequence of elementary operations. Specifically, an $N\times N$ non-unitary $U_{1,2}$ can be expressed as
\begin{equation}
U_{1,2}=VDW=\prod_{n=N}^{n=1}\prod_{m=n-1}^{m=1}V_{mn}D\prod_{i=1}^{i=N}\prod_{j=i+1}^{j=N}W_{ij}. \nonumber
\end{equation}
where $V_{mn}$ and $W_{ij}$ are unitaries acting on a two-dimensional subspace of the system, and $D$ is a diagonal matrix. Each of these operations can be realized using interferometric networks with BDs and wave plates. Importantly, the required number of BDs grows only linearly with $N$, and the parameter count for the wave plates scales quadratically with $N$, ensuring scalability.

Finally, interferometric measurements provide direct access to the eigenenergies in reciprocal space, enabling full reconstruction of the spectrum of arbitrary $N$-level non-Hermitian Hamiltonians.

\end{widetext}

\end{document}